\documentclass{iaus}
\usepackage{graphicx}
\def\lh{l_{\rm h}}
\def\lsoft{l_{\rm s}}

\title[Energy spectra of X-ray QPO] 
{Energy spectra of X-ray quasi-periodic oscillations in accreting black hole
binaries}

\author[P. \.{Z}ycki, M. Sobolewska, A. Nied\'{z}wiecki]   
{P. T. \.{Z}ycki$^1$,  M. A. Sobolewska$^2$ \and A. Nied\'{z}wiecki$^3$}

\affiliation{$^2$Nicolaus Copernicus Astronomical Center, Bartycka 18,
00-716 Warsaw, Poland \break email: ptz@camk.edu.pl \\[\affilskip]
$^2$Department of Physics, University of Durham, South Road, Durham DH1 3LE, 
UK 
\\[\affilskip] 
$^3${\L}\'{o}d\'{z} University, Department of Physics, Pomorska 149/153, 
 90-236 {\L}\'{o}d\'{z}, Poland \break }

\pubyear{2006}
\volume{238}  
\pagerange{001--999}
\date{??? and in revised form ???}
\jname{Black Holes: from Stars to Galaxies -- across the Range of Masses}
\editors{V. Karas \& G. Matt, eds.}
\begin{document}

\maketitle

\begin{abstract}
We investigate the energy dependencies of X-ray quasi-periodic oscillations
in black hole X-ray binaries. We analyze RXTE data on both the low- and
high-frequency QPO. We construct the low-$f$ QPO energy spectra, and 
demonstrate that they do not contain the thermal disk component, even though
the latter is present in the time averaged spectra. The disk thus does not
seem to participate in the oscillations. Moreover the QPO spectra are
harder than the time averaged spectra when the latter are soft, which can
be modeled as a result of modulations occurring in the hot plasma.
The QPO spectra are softer than the time averaged spectra when the latter are 
hard. The absence of the disk component in the QPO spectra is true also
for the high-frequency (hecto-Hz) QPO observed in black hole binaries. We
compute the QPO spectra expected from the model of disk resonances.

\end{abstract}

\firstsection 
\section{Introduction}

Quasi-periodic components of X-ray variability are often seen in power 
density spectra (PDS) from accreting compact objects, 
signifying the existence of ``clocks'' operating on a 
specific time scale, among the broad-band variability. 
The nature of these periodic processes is unknown, but
they are usually associated with various modes of oscillations of standard 
accretion disks present in those systems.

Energy dependencies of the QPO can be an important clue to the origin
of these features, similarly to analyzes of time averaged energy spectra,
which provide information about physical processes of generation of X-rays.
It was noted quite early on into the study of QPO that their r.m.s.\ 
variability seems to increase with energy (review in \cite{vanderKlis2006}). 
Here we report on more detailed investigations  
of QPO energy dependencies in black hole X-ray binaries. 
Firstly, we construct observed low-$f$ (1--10 Hz) QPO energy
spectra in the 2--20 keV range using RXTE data. Secondly, we construct
models of spectral variability of inverse-Compton emission, producing a given
type of QPO spectra. We also analyze data on high-$f$ (hecto-Hz) QPO 
and construct models of energy spectra of these QPO in the model of
disk resonances (\cite{Abramowicz01}).

\section{Low-frequency QPO}

\subsection{Observations}

\begin{figure}
 \includegraphics[height=2.8in]{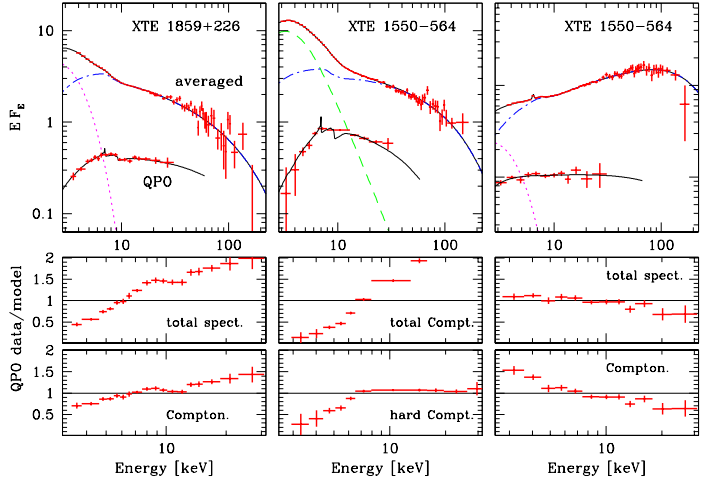}
  \caption{Comparison of time averaged spectra and the QPO spectra (upper
panels. Lower panels show ratios of the QPO spectra to various components
from the time averaged spectra.}
  \label{fig:zycki1}
\end{figure}

We analyzed observations of four sources: GRS~1915+105, XTE~1550-564, 
4U1630-47 and 
XTE~1859+226 performed by RXTE in observing modes providing good energy 
and timing resolution. Details of data reduction and analysis procedures 
are given in \cite{SZ2006}. 
Firstly, we construct time-averaged energy spectra in the 3--200 keV
energy range using PCA and HEXTE data. These are fit with a model
containing two or three continuum components (disk black body, 
comptonization) and the reprocessed component (Fe K$\alpha$ line and the
Compton reflected continuum). Then we construct the QPO energy
spectra: we compute PDS in each energy channel, then fit
it with a model consisting of a broad band continuum (broken
power law or a number of Lorentzians) and one or two narrow Lorentzian
peaks representing the QPO, possibly with a harmonic. The QPO Lorentzians
are then integrated over frequency. This, expressed as variance
(rather than the usual r.m.s./mean) is the QPO energy spectrum. 
The meaning of such a function is that it
would correspond to energy spectrum of a variable component, if it was
a separate spectral component which would be responsible to the QPO.

The results are presented in Fig~\ref{fig:zycki1}. We show both the time 
averaged spectra
and the QPO spectra. We also show the ratios of QPO spectra to the
time averaged models. The main result is the lack of the soft disk component
in the QPO spectra. Presence of this is excluded in all analyzed spectra
at high confidence level. This can be seen for example in the spectra
of XTE~1859+226, where the QPO spectra are much closer to the Comptonized
component only than to total time averaged spectra. Secondly, we analyzed
the slope of the QPO spectra relative to the time averaged spectra.
In soft spectral state the QPO spectra are clearly harder than the time 
averaged spectra (Fig~\ref{fig:zycki1}a), while the opposite is true for 
the hard state (Fig~\ref{fig:zycki1}c). During the two observations of 
4U1630-47 the time
averaged spectra are intermediate in slope between values typical
for the soft and hard states ($\Gamma \approx 2$). We find that the QPO
spectra are similar to the time averaged spectra in these two observations.

\subsection{Models}

In an attempt to understand the results presented in the previous Section,
we performed simulations of spectral variability of inverse-Compton
component. Such spectra are described by three main parameters: the plasma 
heating
rate, $\lh$, the flux of cooling soft photons, $\lsoft$, and the temperature
of the input soft photons, $T_0$. Irrespectively on the physical mechanism
of QPO production, it is one of the three parameters that has to be modulated
if the observed QPO corresponds to modulation of intrinsic luminosity
(other possibility is a periodic modulation of geometry affecting, e.g.,
relativistic effect influencing the observed luminosity; see below).

\begin{figure}
 \includegraphics[height=1.9in]{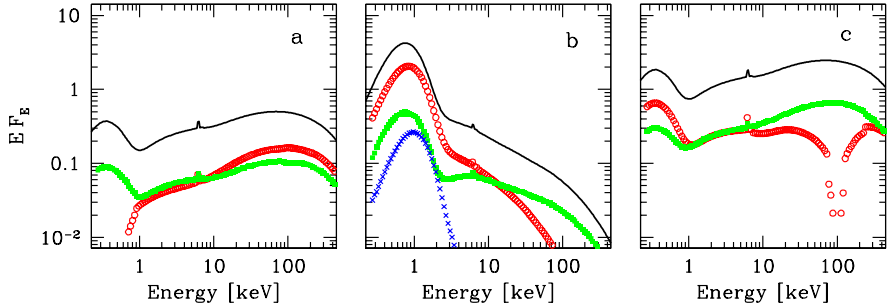}
  \caption{Model predictions for the QPO energy spectra (red circles)
  in various scenarios for modulations: a) heating, b) cooling, c)
  covering factor of cold matter. Green squares are the broad band noise
spectra, blue stars: QPO harmonic, solid line shows time averaged spectra.}
\label{fig:zycki2}
\end{figure}

The slope of the time averaged Comptonized
spectrum depends (almost) only on the ratio $\lh/\lsoft$, however the mode
of spectral variability depends on which parameter is modulated. 
Details of this investigation are given in \cite{ZS2005}.
To summarize it: modulation of the heating rate produces variability spectrum
with largest amplitude of variability at high energies, which means that
the variability spectrum is {\em harder\/} than the time averaged spectrum. 
Modulation
of the cooling rate or the soft photons temperature produces variability
spectrum {\em softer\/} than the time averaged spectrum. We considered
also modulation of the amplitude of the reprocessed component, which was
assumed to relate the cooling and heating rate (see, e.g., 
\cite{PF99}. In this case we also find that the QPO are softer
than the time averaged spectra, but, additionally, a very strong Fe K$\alpha$
line is seen in the QPO spectrum.

Summarizing, properties of low-$f$ QPO spectra in the soft state  strongly 
suggest that these QPO are generated as a result of modulations in the hot 
plasma rather than in the standard cold accretion disk. This is contrary to
most theoretical ideas where oscillations of the cold disk are thought to be
the primary driver. However, the physical mechanism does seem to require
both accreting plasma phases to work, since these QPO are associated
with changes of the geometry of the system. It is also interesting to note,
that long-time scale variability of Cyg X-1 in the soft state also seems to be
driven by instabilities in the hot Comptonizing plasma (\cite{Zdziarski2002}).

\section{High-frequency QPO}

\begin{figure}
 \includegraphics[height=1.9in]{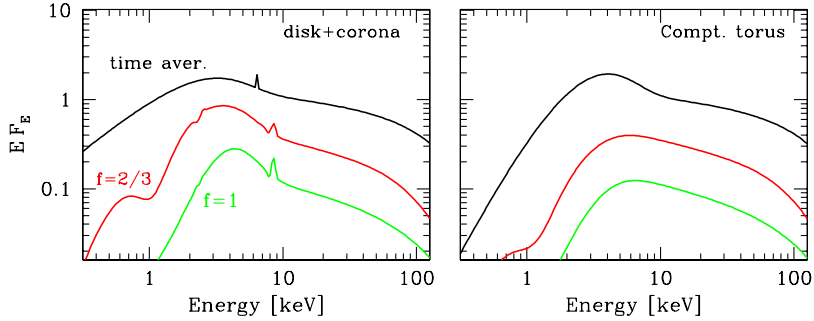}
  \caption{QPO spectra from the model of epicyclic oscillations, with
 relativistic effects modulating the X-ray flux, for $a=0.998$ and inclination
$\mu_{\rm obs}=0.3$.  Left: geometry of accretion 
 disk with an active corona produces strong disk component in the QPO spectra.
 Right: hot oscillating torus produces QPO spectrum resembling a comptonization 
component.}\label{fig:zycki3}
\end{figure}

Data on high-frequency QPO in black hole X-ray binaries are of much poorer
quality than data for the low-$f$ QPO. It is nevertheless clear that they share
at least one characteristics with their low-$f$ counterparts, namely the soft
disk component is absent in the QPO spectra. More detailed energy dependencies
(i.e., the spectral slope) are not possible to determine with current data.
The fact that pairs of QPO with 3:2 frequency ratio are sometimes 
observed, stimulated development of models invoking non-linear resonances
between various modes of oscillations of accretion disks. In particular,
\cite{Abramowicz01} postulate that the relevant modes are the vertical
and radial epicyclic motions. Importantly, they suggest also that the X-ray
flux modulation occurs due to changes of geometry and, consequently, changes 
in the strength of the relativistic effects (Doppler and gravitational 
shifts; light bending), with the primary emission being constant. This is
thus the only current model making definite predictions for the more detailed
properties of the QPO, for example, their energy spectra. We used the code of 
\cite{ZN2005} for photon transfer in Kerr metric and modified it 
to implement the source motion corresponding to disk epicyclic oscillations 
(details are in \.{Z}ycki et al., in preparation). 
We have performed computations of the relativistic effects expected from 
such disks at radii corresponding to 3:2 resonance between the oscillation
frequencies, and we have computed energy spectra of QPO expected
in this model. 
They are presented in Fig.~\ref{fig:zycki3} in two variants of geometry:
(1) accretion disk extending to the last stable orbit with an active X-ray 
corona and (2) a truncated accretion disk with inner hot Comptonizing flow. 
In the first scenario the QPO spectrum contains a strong disk component, since
the disk emission is modified by the relativistic effects. In the second 
scenario the disk is, by assumption, not present at the QPO radius and, as 
a consequence, only the Comptonized component is modulated. The Fe K$\alpha$
line is also assumed to be produced at the QPO radius (in the accretion
disk with a corona geometry) and it does appear
in the QPO spectrum, at positions corresponding to maximum redshift expected
from the oscillatory motion.

\begin{acknowledgments}
This work was partially supported by grant 2P03D01225 from Polish Ministry of
Science (MNIiI).
\end{acknowledgments}


\begin{thebibliography}{}

\bibitem[Abramowicz \& Klu\'{z}niak (2001)]{Abramowicz01}
     {Abramowicz, M. A. \& Klu\'{z}niak, W.} 2001
     \textit{A\&A} 374, L19

\bibitem[Poutanen \& Fabian (1999)]{PF99}
     {Poutanen, J. \& Fabian, A. C.} 1999
     \textit{MNRAS} 306, L31

\bibitem[Sobolewska \& \.{Z}ycki (2006)]{SZ2006}
     {Sobolewska  M. A. \& \.{Z}ycki, P. T.} 2006
     \textit{MNRAS} 370, 405

\bibitem[van der Klis (2006)]{vanderKlis2006}
     {van der Klis, M.} 2006,
     in: W. H. G. Lewin \& M. van der Klis (eds.), 
     \textit{Compact Stellar X-ray Sources} (Cambridge: CUP), p.\ 39 

\bibitem[Zdziarski et al. (2002)]{Zdziarski2002}
     {Zdziarski, A. A., Poutanen, J., Paciesas, W. S., Wen, L.} 2002
     \textit{ApJ} 578, 357

\bibitem[\.{Z}ycki \& Nied\'{z}wiecki (2005)]{ZN2005}
     {\.{Z}ycki, P. T. \& Nied\'{z}wiecki A.} 2005
     \textit{MNRAS} 359, 308

\bibitem[\.{Z}ycki \& Sobolewska (2005)]{ZS2005}
     {\.{Z}ycki, P. T. \& Sobolewska  M. A. } 2005
     \textit{MNRAS} 364, 891


\end{thebibliography}
\end{document}